\documentclass[%
 reprint,
 amsmath,amssymb,
 aps,
]{revtex4-2}

\usepackage{graphicx}
\usepackage{dcolumn}
\usepackage{bm}
\usepackage{xcolor,soul,framed} 
\usepackage{subfigure}

\begin{document}


\title{Analysis of Metallic Space-Time Gratings using Lorentz Transformations}


\

\author{Antonio Alex-Amor}
\affiliation{
Department of Information Technology, 
Universidad San Pablo-CEU, CEU Universities,  Campus Montepríncipe, 28668 Boadilla del Monte (Madrid), Spain
}%
\author{Carlos Molero}
\affiliation{%
Department of Signal Theory, Telematics and Communications, Universidad de Granada, 18071 Granada, Spain
}%
\author{Mário G. Silveirinha}
\affiliation{
Instituto Superior Técnico and Instituto de Telecomunica\c{c}oes, University of Lisbon, 1049-001 Lisboa, Portugal 
}%

\begin{abstract}
This paper presents an analytical framework for the study of scattering and diffraction phenomena in spacetime-modulated metallic gratings. Using a Lorentz transformation, it is shown that a particular class of spacetime-modulated gratings behave effectively as moving media. We take advantage of this property to derive a closed analytical solution for the wave scattering problem. In particular, using our formalism it is possible to avoid spacetime Floquet-Bloch expansions, as the solution of the problem in the original laboratory frame (grating parameters are periodic in space and time) is directly linked to a co-moving frame where the metallic grating is time-invariant (grating parameters are periodic only in space). In this way, we identify a fundamental connection between moving metallic gratings and spacetime-modulated metamaterials, and exploit this link to study the nonreciprocal response of the structure. 
Some limitations and difficulties of the alternative non-relativistic Galilean approach are discussed and the benefits of the Lorentz approach are highlighted. 
Finally, some analytical results are presented in order to validate the formalism. The results include scenarios involving TM($p$) and TE($s$) normal and oblique incidence, even beyond the onset of the diffraction regime. Furthermore, we show how the synthetic Fresnel drag can tailor the Goos-Hänchen effect and create a specular point shifted towards the direction of the synthetic motion, independent of the sign of the incidence angle. 
\end{abstract}

\maketitle


\section{\label{sec:introduction}Introduction}

Diffraction gratings enable peculiar and fascinating optical behaviors, which were observed already in the 18th century  \cite{Rittenhouse1786}.The scattering by a space grating is characterized by several resonant features, the most remarkable of these being the so-called Wood's anomaly \cite{Wood1902}. Nowadays, diffraction gratings are useful in a wide range of electromagnetic systems \cite{Palmer2005}, which typically exploit their dispersive properties and frequency dependent response. 
Furthermore, the scattering of light by space gratings is important in spectroscopy, and the development of this field has benefited from the theoretical and experimental progress in optical diffraction gratings during the last and the present century \cite{Sokolova2015}. 

Historically, diffraction-gratings research was mainly focused in  optics and in the angular separation of polychromatic light into different monochromatic components. The accumulated knowledge in this field stimulated the  study of applications in other parts of the electromagnetic spectrum \cite{Lord1957, Hibbins2004, Alex3D_2022},  in acoustic systems \cite{Greenwood2006} and others, for particle detection and measurement \cite{Grunzweig2008}. 

The effect of motion in the wave diffraction by a grating has also been object study for several decades in different contexts, for example in neutron-beam scattering \cite{kowalski2006, Frank2003}. Moreover, moving gratings have been   
employed for medical diagnosis of the human vision \cite{Tolhurst1973} and studies of the vision of animals \cite{Lee1981}.
Such systems exhibit spacetime properties, as the position of the grating elements changes in time \cite{Kelly1982, Mazor2020}.

The grating motion can tailor in unique ways the scattering and the diffraction of electromagnetic waves.
For example, the motion of the grating in the direction of propagation of the incoming wave leads to Doppler shifts in the frequency of the scattered waves  \cite{Dossou2016}.
Spatial non-reciprocity is another interesting property provided by a translational motion \cite{Bahabad2014}. The nonreciprocity manifests itself in an asymmetry of the diffraction patterns originated by the interaction between the grating and incident waves arriving from complementary directions of space, linked by Snell's laws. Different from reciprocal systems, the generated diffraction patterns may depend on the direction of arrival of the wave. A related effect was recently studied in the context of space-temporal modulated systems \cite{spacetime_caloz1, spacetime_caloz2}.

Moving gratings may be regarded as a particular instance of time-variant systems. This is because the environment  wherein the wave propagates changes in time due to the physical motion.
Preliminary studies on moving (dielectric) media and time-variant systems date back to the last century \cite{Yeh1966}. In recent years,  the interaction of waves with more general time-variant systems, e.g., in systems with materials or components whose response is electrically modulated in time, has been a topic of intense scientific research \cite{Hadad2015, Taravati_magazine2020, Taravati2022, AimingPacheco, Pacheco-Pena2022, Galiffi2022, Mazor2_2019}. In general, the physical response of a spacetime modulated system is $not$ equivalent to that of a moving dielectric material, even for the simplest case of a ``travelling wave'' modulation of the material permittivity such that $\varepsilon \left( {x,t} \right) = \varepsilon \left( {x - v\,t} \right)$, with $v$ the modulation speed. In fact, the electromagnetic response provided by a travelling-wave modulation differs in many ways from the response of a moving dielectric material \cite{Galiffi2022, Yeh1966}. Curiously, in the homogenization limit it is possible to link precisely the response of some  spacetime crystals with the response of an equivalent moving medium \cite{Huidobro2019, Huidobro2021, Kreiczer2021}. However, it is underlined that such equivalence is applicable only in the effective medium limit and for nondispersive isotropic crystals.

In this article, we study a class of time-variant platforms which can be equivalently regarded as either ``moving systems'' or ''space-time modulated systems''. Specifically, we consider a spacetime grating formed by thin perfect electric conducting (PEC) plates. As it is well known, the PEC boundary condition is unaffected by a Lorentz boost when the relative velocity is parallel to the PEC surface (see Appendix A for further details). We take advantage of this property to show that a spacetime modulated grating formed by thin metallic plates is exactly equivalent to the corresponding moving metallic grating. Using this property and Lorentz transformations, it becomes possible to connect the solution of a scattering problem in the original frame where the grating is time-varying, with the solution of an equivalent scattering problem in a co-moving frame where the grating is time invariant.


Our approach is particularly interesting due to the scarcity of commercial full-wave solutions that are able to model spacetime systems. The few commercial time-domain solutions that can include time modulations operate with important limitations, although some studies have been already aided by their use \cite{AimingPacheco, Garg2022}. Other specific (not general) alternatives try to solve the scattering problem by using commercially-available frequency-domain solvers. For instance, the method developed in \cite{Sievenpiper2019} divides the time modulation into discrete time steps, which are individually simulated and then synthesized to compose the complete response. These alternatives, although ingenious, are inefficient and typically restricted to some particular regime of operation of the considered spatiotemporal system. As commercial solutions are not well developed yet, the analysis of spacetime metasurfaces is typically done in terms of cumbersome spacetime Bloch modal expansions  \cite{Taravati2019, Zurita2009, Wu2020}. Other numerical approaches include the Method of Moments (MoM) \cite{Bass2022}, circuital interpretations \cite{Wang2020}, and others \cite{Ptitcyn2019, Tiukuvaara2021}. 
In several cases, validations of the models are done by comparison with the results provided by home-made finite-difference in the time-domain (FDTD) codes \cite{Taravati2019, Stewart2018, Vahabzadeh2018, Timevarying2022, Itoh1989, FDTDCaloz2023}.

The characterization of the dynamical response and scattering parameters (reflection and transmission coefficients) in a spacetime grating is a nontrivial task. Some recent proposals exploit the spacetime periodicity of the structure by means of ($N+1$) Floquet-Bloch modal expansions, with $N$ the space dimension. Although these are interesting approaches, their mathematical and computational complexity increases notably with $N$. Here, we will exploit an interesting alternative which relies on the use of circuital models that have previously been introduced for time-invariant (static) gratings \cite{Medina2010, FloquetCircuit1, FloquetCircuit2}. In such a framework, the grating is modelled by a homogenized impedance determined by the grating geometry. The most sophisticated circuit models can take into account the effect of grating lobes \cite{FloquetCircuit1, FloquetCircuit2}.


Special relativity theory establishes the equivalence between physical phenomena witnessed by observers moving at different speeds via Lorentz transformations \cite{lorentz_transformations}. Gratings moving with a constant velocity can always be studied in a coordinate system where they are at rest (co-moving frame). Since the circuit model of a static grating is known,  with the help of a Lorentz transformation it is possible to find the proper equivalent impedance in another inertial frame. This work develops equivalent circuits for moving gratings by combining Lorentz transformation and circuit models derived in static scenarios. Using our approach, we are able to study in a simple way how the nonreciprocity provided by the time-modulation (or translational motion) can tailor the wave scattering, the diffraction resonances, the Goos-Hänchen shift \cite{Lakhtakia2004, Fan2018}, and other phenomena.





The manuscript is organized as follows. Section II presents brief overview of Lorentz transformations and the analytical framework used in our study.
The equivalence between moving and spacetime-modulated metallic gratings is established. Using that result, we obtain an equivalent circuit for the time-modulated grating. Section III presents analytical and numerical results that illustrate the application of our theory. In particular, we use our circuit model to characterize the wave diffraction by a space-time modulated gratings under normal and oblique incidence conditions. Furthermore, we show how the modulation speed can tailor the sign and strength of the Goos-Hänchen shift. The main conclusions are drawn in Section IV.

\section{\label{sec:analytical}Analytical Framework}

\begin{figure}[!t]
	\centering
	\subfigure{ \hspace{-0.8cm}
		\includegraphics[width=0.54\textwidth]{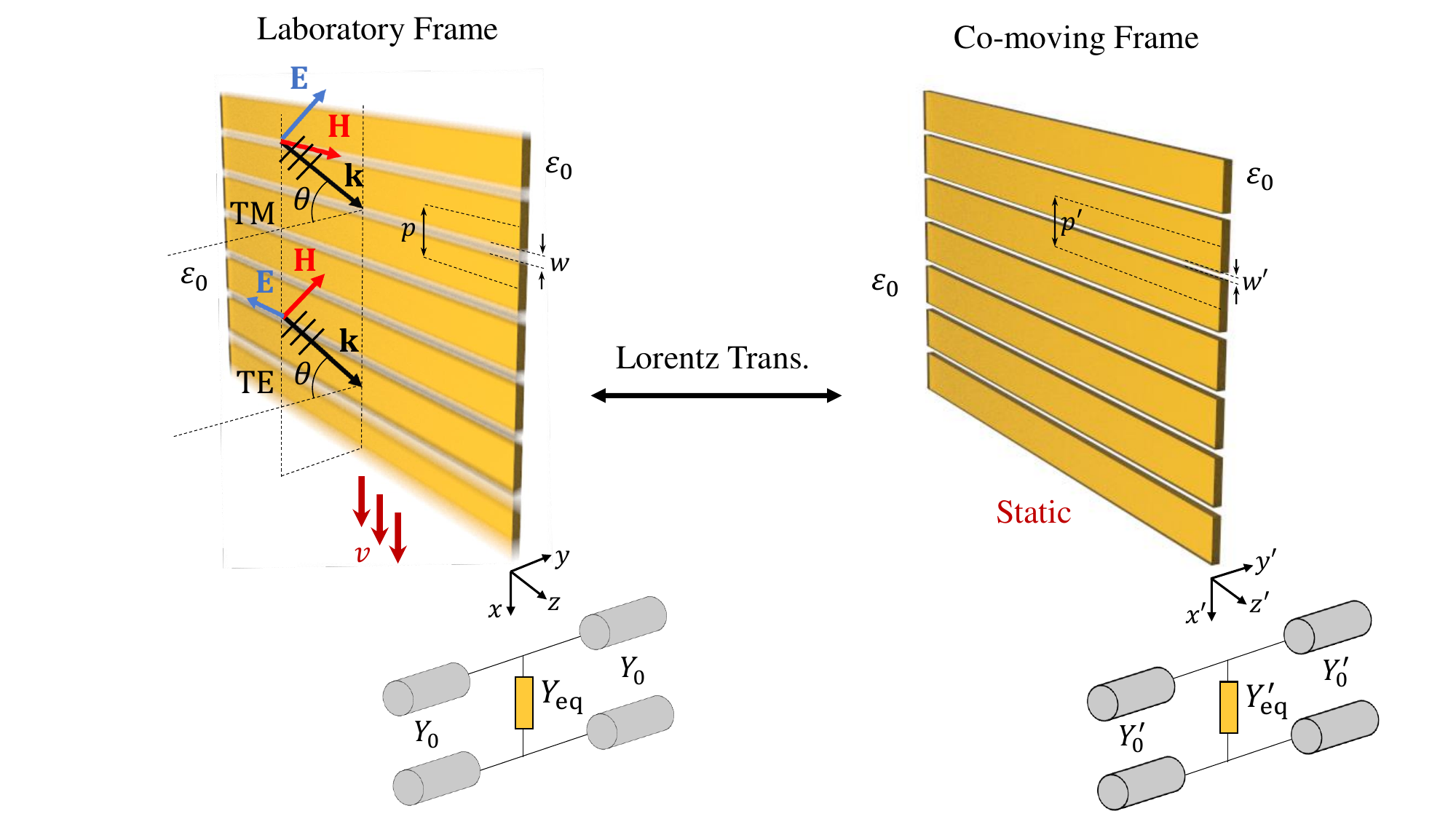}} \vspace{-0.5cm}\\
	\caption{\small Moving grating in the original space (laboratory frame) and its static equivalent in the transformed space (co-moving frame). The wave scattering in the laboratory and co-moving frames is modelled using equivalent circuits. The parameters of the equivalent circuits are related via Lorentz transformations. } 
	\label{fig1}
\end{figure}

The scenario under analysis is sketched in the left panel of Figure \ref{fig1}. A monochromatic plane wave of angular frequency $\omega$  impinges obliquely, with incident angle $\theta$, on a diffraction metallic (PEC) grating that ``moves'' \emph{uniformly} with velocity $v$ along the $x$ axis. The motion of the grating may be either synthetic, i.e., created by an electric modulation, or  actual physical motion. Below, we shall show that for thin PEC gratings the two problems are equivalent. The diffraction grating is formed by a periodic arrangement (period $p$) of metallic strips separated by slits of width $w$. The strips are infinitely extended along the $z$ direction. The metallic plates thickness is infinitesimal and they are surrounded by air. 
Two different polarizations are considered for the incident plane wave: transverse magnetic, TM ($p$), and transverse electric, TE ($s$).

\subsection{Equivalence between moving media and the spacetime modulation for the metallic gratings}

As discussed in the introduction, the study of time-varying diffraction gratings is a complex task. On the other hand, the analysis of their  1D \cite{ FloquetCircuit1, FloquetCircuit2, FloquetCircuit3} and 2D \cite{FloquetCircuit0, FloquetCircuit_2D, FloquetCircuit_2D1, FloquetCircuit_2D2} time-invariant (static) counterparts is a well-established topic in microwave and photonics communities. From a circuit standpoint, a static grating may be modeled as an equivalent admittance/impedance (see Appendix B for details) placed in between two transmission lines. The transmission lines describe the propagation in input and output semi-infinite media \cite{Mesa2018}. 

The wealth of knowledge accumulated over decades of research on static gratings provides a valuable foundation for investigations into the properties of moving gratings. The obvious idea is to switch to the so-called \emph{co-moving frame}, where the observer moves with the same velocity as the grating, see the right-hand side panel in Figure \ref{fig1}. The grating response is independent of time (static) in the co-moving frame. 

There is however a practical difficulty in using the solution of the time-invariant canonical problem in the co-moving frame coordinates. It is related to the fact that the constitutive relations of the material $depend$ on the inertial frame. For example, a standard isotropic dielectric is seen in a generic inertial frame as a bianisotropic material. Due to this reason, a spacetime-modulated grating [$\varepsilon \left( {x,t} \right) = \varepsilon \left( {x - v\,t} \right)$] has an electromagnetic response rather different from that of the corresponding moving grating \cite{lorentz_transformations, Galiffi2022, Prudencio2023}. Another difficulty is related to material dispersion. In fact, under a Lorentz transformation the frequency and the wave vector are mixed through a Doppler transformation, giving rise to both frequency and spatial dispersion.

The previous discussion highlights that switching to the co-moving frame coordinates is not always beneficial (particularly in case of spacetime modulated systems), as it can originate both a bianisotropic coupling and a complex spatially dispersive response, i.e., lead to a physical scenario very different from the standard static grating. 

Consider however a situation in which all the materials of the system have a response that stays invariant under the relevant Lorentz boost. For example, air (vacuum) is described by the same constitutive relations in all inertial reference frames, and thereby it provides a (trivial) example of a material whose response is frame independent. Less trivial, it can be shown that any nondispersive material ($\varepsilon$, $\mu$) with refractive index identical to $n=1$ has exactly the same property \cite{Prudencio2023}. When the constitutive relations are frame independent, the structure of the dynamical equations is unaffected by the Lorentz transformation, apart from the coordinate transformation in the material parameters. For such a class of systems the spacetime modulation is indistinguishable from a translational physical motion, as switching to the co-moving frame is formally equivalent to setting the modulation speed $v$ identical to zero.

An important example of a material (boundary) that is Lorentz-invariant is the PEC boundary condition. Provided the velocity $\bf{v}$ that determines the Lorentz boost is parallel to relevant PEC boundary, then the response is frame independent (see Appendix A). This means that the tangential electrical field vanishes at the boundary interface, independent of the considered frame. In contrast, when $\bf{v}$ is perpendicular to the relevant boundary, the response may change and the Lorentz invariance is lost.

As an application, consider a metallic grating formed by PEC strips standing in air, with the relevant modulation speed parallel to the interface (Figure \ref{fig1}). Provided the strips are infinitesimally thin, all the relevant PEC interfaces are parallel to $\bf{v}$ and thus the constitutive relations that describe the response of the entire structure (air and PEC strips) are frame independent. Thus, this type of grating provides an interesting example of a spacetime modulated system with a response that is indistinguishable from that of the corresponding moving grating.

In the following subsections, we use a Lorentz transformation to derive the equivalent circuit for the spacetime modulated grating (in the original laboratory frame) from the equivalent circuit in the co-moving frame \cite{lorentz_transformations}.
We will establish a relation between the equivalent admittances in the laboratory ($Y_\mathrm{eq}$) and co-moving ($Y'_\mathrm{eq}$) frames via the original and transformed electromagnetic fields.
It will be shown that the  circuit topology in the laboratory and co-moving frames is the identical, although the value of equivalent admittance changes. 

\subsection{Lorentz transformations}
As previously stated, Lorentz transformations may be used to link the electrodynamics of the problem in the original domain (laboratory frame) with the electrodynamics in a transformed domain (co-moving frame, prime notation $'$) where the grating is static. Next, we present a brief overview of the basic concepts associated with Lorentz transformations.

\subsubsection{Spacetime coordinates}
Assuming that the metallic grating ``moves'' with uniform speed $v$ along the $x$-direction, the spacetime coordinates in the co-moving
frame ($x',y', z', t'$; primed coordinates) are related to the coordinates of the laboratory frame \linebreak ($x, y, z, t$; unprimed coordinates) through the following expressions \cite{lorentz_transformations}
\begin{equation}
	x' = \gamma (x - vt), \quad y' = y,\quad z' = z, \quad t' = \gamma \Big(t - x \frac{v}{c^2} \Big)	
\end{equation}
where $\gamma = 1/\sqrt{1 - v^2/c^2}$ is the Lorentz factor.

\subsubsection{Spectral transformations}
The phase of a wave is  observer independent. Thereby, the frequency and the wave vector of a wave in two reference frames must satisfy $\mathbf{k} \cdot \mathbf{r} -\omega t = \mathbf{k}' \cdot \mathbf{r}' -\omega' t' $. Thus, the frequency ($\omega \rightarrow \omega'$) and wavenumber ($k_x \rightarrow k_x'$) transformations (relativistic Doppler transformation)  are of the form \cite{jackson_electro, uniform_caloz, paloma_spacetime}
\begin{equation} \label{lorentz_omega}
\omega' = \gamma (\omega - v k_x)
\end{equation}
\begin{equation} \label{lorentz_kx}
k_x' = \gamma \Big(k_x - \frac{v}{c^2} \omega \Big), \quad k_y' = k_y, \quad k_z' = k_z
\end{equation}

\subsubsection{Field transformations}
For the scenario depicted in Figure \ref{fig1}, the fields in the co-moving frame are related to fields in the laboratory frame as follows \cite{lorentz_transformations}
\begin{equation} \label{fields1}
E_x' = E_x, \qquad B_x' = B_x
\end{equation}
\begin{equation}
E_y' = \gamma \Big(E_y - vB_z \Big), \qquad B_y' = \gamma \Big(B_y + \frac{v}{c^2}E_z\Big)
\end{equation}
\begin{equation} \label{fields3}
E_z' = \gamma \Big(E_z + vB_y\Big), \qquad B_z' = \gamma \Big(B_z - \frac{v}{c^2}E_y\Big)
\end{equation}
As it can be seen, the field components along the direction of motion ($x$-axis) are identical in the two frames, while the field components orthogonal to the direction of motion change under a Lorentz transformation. Additionally, as already discussed in Sect. II.A,  it can be readily inferred from the field equations \eqref{fields1}-\eqref{fields3} that the synthetic motion of the grating may have important implications on the material response (constitutive parameters) in the co-moving frame.

In fact, conventional dielectrics and isotropic media in the (unprimed) laboratory frame transform into bianisotropic materials in the (primed) co-moving frame \cite{spacetime_caloz1, Vehmas_2014, Prudencio2023}. This can be understood by noting that an electric dipole in a rest frame is seen as a superposition of electric and magnetic dipoles by a moving observer due to the so-called Röntgen current, leading to a bianisotropic magnetodielectric coupling \cite{Wilkens94}. 


Nonetheless, the vacuum (and thereby also air) and perfect electric conductors (PEC) are special media whose constitutive parameters \emph{remain invariant} under Lorentz transformations; namely, these media have precisely the same electromagnetic response  in both laboratory and co-moving frames. For the vacuum case, this property is a trivial consequence of the invariance of the free-space Maxwell equations under Lorentz transformations. Strictly speaking, the PEC response is frame independent only if the direction of motion is parallel to the PEC walls. Thus, we require that the thickness of the PEC plates is vanishingly thin, so that the response of the sidewalls can be disregarded in the analysis.
Thus, in this study we suppose that the input and output regions in Figure \ref{fig1} are air, while the metals are thin perfect electric conductors (PEC) plates. Under these circumstances, the analysis of the problem is greatly simplified as bianisotropic material response is avoided in the co-moving frame.

It is important to note that even though the response of the metal plates is identical in the two frames, the Lorentz transformation of coordinates leads to the so-called Lorentz-FitzGerald contraction of all the lengths along the direction of motion \cite{jackson_electro}. Specifically, the lattice period and the air slit widths in the laboratory frame are shorter than in the co-moving frame: $p=p'/\gamma$ and $w=w'/\gamma$. Thus, from the point of view of the laboratory frame, the original geometry suffers a dilation along the $x$-direction after the coordinate transformation.

\subsection{Circuit model}

 The problem shown in Figure \ref{fig1} can be analyzed with an equivalent circuit model. The input ($i=1$) and output ($i=2$) air regions can be modeled as semi-infinite transmission lines with wave admittances \linebreak $Y_0^\mathrm{(1)} = Y_0^\mathrm{(2)} = Y_0$. In the homogenization limit, the infinitesimally-thin moving grating may be represented as a shunt equivalent admittance $Y_\mathrm{eq}$ or surface impedance of \textit{unknown} value \cite{Mazor2_2019}. 
 
 Once the equivalent admittance $Y_\mathrm{eq}$ that models the moving grating has been determined (see the next subsections), one can find the reflection $R$ and transmission $T$ coefficients of the structure as
 \begin{equation} \label{R}
     R = \frac{Y_{0}^{(1)} - Y_{0}^{(2)} - Y_\mathrm{eq}}{Y_{0}^{(1)} + Y_{0}^{(2)} + Y_\mathrm{eq}}
 \end{equation}
\begin{equation}
    T = 1 + R
\end{equation}
In our case, the input and output (free-space) regions are identical ($Y_{0}^{(1)} = Y_{0}^{(2)} = Y_{0}$), so that Eq. \eqref{R} can be simplified to
\begin{equation} \label{R2}
    R = \frac{-Y_\mathrm{eq}}{2 Y_{0} + Y_\mathrm{eq}}
\end{equation}
These equations are obtained from the equivalent circuit model of the spacetime grating (see Figure \ref{fig1}). The value of $Y_\mathrm{eq}$ is dependent on the considered excitation: TM or TE. In the next subsections, we use a Lorentz transformation to link the equivalent admittance of the time-independent grating ($Y'_\mathrm{eq}$, evaluated in the co-moving frame) with the equivalent admittance of the spacetime grating $Y_\mathrm{eq}$.

\subsubsection{TM ($p$) Incidence ($H_x = H_y = E_z = 0$)}
Let us first consider that a TM-polarized plane wave impinges on the spacetime grating of Figure \ref{fig1}. For TM incidence, the magnetic field is oriented as  $\mathbf{H} =  H_z \, \hat{\mathbf{z}}$. The field components parallel to the metallic interfaces are along $x$ and $z$ directions. Thus, the objective here is to relate the $x$ and $z$ components of the fields, at the interface $y=0$, in the laboratory frame. 

The primed equivalent admittance in the co-moving frame, $Y_\mathrm{eq}'$, relates the tangential electric field with the discontinuous magnetic field according to \linebreak $\mathbf{\hat{n}}' \times [\mathbf{H}'] = Y_\mathrm{eq}' \, \mathbf{E}'_t$. The rectangular brackets $[\,\cdot\,]$ stand for the jump discontinuity of the field at the interface. Therefore, the equivalent admittance in the co-moving frame is defined in such a way that:
\begin{equation}
	Y_\mathrm{eq}' = \frac{H_z'|_{y=0^+} - H_z'|_{y=0^-}}{E_x'}, \quad \textrm{TM pol.}
\end{equation}
Using Eqs. \eqref{fields1}-\eqref{fields3} to relate the fields in the laboratory and co-moving frames, it is found that
\begin{equation}
     Y_\mathrm{eq}'= 
	\frac{\gamma \Big[(H_z - v\varepsilon_0E_y)|_{y=0^+} - (H_z - v\varepsilon_0E_y)|_{y=0^-} \Big]}{E_x}
\end{equation}
Reorganizing the different terms, one obtains the following boundary condition in the laboratory frame:
\begin{equation} \label{hom}
\frac{Y_\mathrm{eq}'}{\gamma} = 
\frac{H_z|_{y=0^+} - H_z|_{y=0^-}}{E_x} - v\varepsilon_0 \frac{ E_y|_{y=0^+} - E_y|_{y=0^-}}{E_x}
\end{equation}
Besides the above constraint, one should also enforce that $E_x$ is continuous at the two grating sides, which is a consequence of $E_x = E'_x $. 

In order to further simplify Eq. \eqref{hom}, next we write  the jump discontinuity of the normal component of the electric field,  $E_y|_{y=0^+} - E_y|_{y=0^-}$, in terms of the tangential magnetic field ($H_z$).
This can be done by using Maxwell's equations, specifically, $\nabla \times \mathbf{H} = j \omega \varepsilon_0 \mathbf{E}$, which yields:
\begin{equation}
E_y|_{y=0^+} - E_y|_{y=0^-} = \frac{1}{-j \omega \varepsilon_0}\, \partial_x \left(H_z|_{y=0^+} - H_z|_{y=0^-} \right)
\end{equation}
where $\partial_x$ refers to the partial derivative with respect to $x$. For a dependence on $x$ and $t$ of the type ${e^{j\omega t}}{e^{ - j{k_x}x}}$, one can use $\partial_x = -jk_x$. Thus, it follows that
\begin{equation}
E_y|_{y=0^+} - E_y|_{y=0^-} = \frac{k_x}{\omega \varepsilon_0} \left(H_z|_{y=0^+} - H_z|_{y=0^-} \right).
\end{equation}
Substituting the above equation into \eqref{hom} leads to
\begin{equation} \label{hom3}
\frac{Y_\mathrm{eq}'}{\gamma} = 
\frac{H_z|_{y=0^+} - H_z|_{y=0^-}}{E_x} - \frac{v k_x }{\omega}\, \frac{H_z|_{y=0^+} - H_z|_{y=0^-}}{E_x}.
\end{equation}
The equivalent admittance in the laboratory frame is defined in the standard way as \linebreak $Y_\mathrm{eq} = (H_z|_{y=0^+} - H_z|_{y=0^-})/E_x$. Thus, the previous analysis shows that  the admittances in the laboratory and co-moving frames are linked as:
\begin{equation} \label{hom4}
Y_\mathrm{eq}(\omega, \mathbf{k}) = \frac{Y_\mathrm{eq}'(\omega', \mathbf{k}')}{\gamma (1 - \frac{vk_x}{\omega})}, \quad \textrm{TM pol.}
\end{equation}
In \eqref{hom4}, the transverse wavenumber $k_x$ is related to the incident angle $\theta$ as $k_x = k_0 \sin \theta$, with $k_0 = \omega / c$. Therefore, \eqref{hom4} may be equivalently rewritten as
\begin{equation} \label{hom5}
    Y_\mathrm{eq}(\omega, \mathbf{k}) = \frac{Y_\mathrm{eq}'(\omega', \mathbf{k}')}{\gamma (1 - \frac{v }{c}\sin \theta)}, \quad \textrm{TM pol.}
\end{equation}
Detailed information on how to calculate the equivalent admittance in the co-moving frame, $Y_\mathrm{eq}'$ for a static grating of thin PEC plates, is given in Appendix \ref{sec:appendixa}.

By examining Eq. \eqref{hom5}, one can draw several conclusions. The most important observation is that the equivalent circuit in the laboratory frame has exactly the  \textit{same structure} as the one in the (static) co-moving  frame. The circuit is formed by two semi-infinite transmission lines (related to the input and output air regions) and by an equivalent admittance $Y_\mathrm{eq}$ that models the moving grating. This is a remarkable feature, which allows to simplify enormously the analysis of scattering phenomena in spacetime gratings. It is important to highlight that due to synthetic motion the frequency and wave vector in the argument of $Y_\mathrm{eq}'(\omega', \mathbf{k}')$ are affected by a Doppler shift. 
As to the term in the denominator of \eqref{hom5}, one can infer that it is typically negligible at low modulation speeds under a paraxial approximation. In fact, the term $v\sin (\theta)/c $ is is precisely zero for normal incidence ($\theta = 0$).

\subsubsection{TE ($s$) Incidence ($E_x = E_y = H_z = 0$)}
For TE-polarization, the electric field is oriented as $\mathbf{E} = -E_z \, \hat{\mathbf{z}}$. In this case, the equivalent admittance in the co-moving frame is defined by
\begin{equation}
Y_\mathrm{eq}' = \frac{H_x'|_{y=0^+} - H_x'|_{y=0^-}}{E_z'}, \quad \textrm{TE pol.}
\end{equation}
By repeating the same steps as for the TM case, it can be demonstrated that for TE-polarization $Y_\mathrm{eq}$ and $Y'_\mathrm{eq}$ are related as
\begin{equation} \label{homTE}
    Y_\mathrm{eq}(\omega, \mathbf{k}) =  \gamma \left(1 - \frac{v }{c}\sin \theta \right)\, Y_\mathrm{eq}'(\omega', \mathbf{k}'), \quad \textrm{TE pol.}
\end{equation}
Note that in this case the term $1 - v\sin (\theta)/c $ appears as a multiplication factor, different from the TM polarization where it appears as a dividing factor. It is also worth pointing out that the tangential field $E_z$ is continuous in the laboratory frame.


\subsection{Diffraction resonances}

Similarly to static gratings, spacetime gratings can excite higher diffraction orders, which correspond to different diffraction beams. Heuristically, the onset of the different diffraction orders is expected to be Doppler-shifted as a consequence of the synthetic motion \cite{Askne1971}.  

As is well-known, for static gratings the diffraction resonances appear when the wave admittance of the considered order diverges (vanishes) for TM (TE) polarization. Hence, the onset of diffraction in the co-moving frame is determined by  $Y'_n \rightarrow \infty$ (TM) and $Y'_n \rightarrow 0$ (TE), with $Y'_n$ the free-space wave admittance of the $n$-th diffraction order. In either case, the free-space longitudinal wavenumber must vanish,
\begin{equation}
    k'_{yn} = \sqrt{k_0^{'2} - k_{xn}^{'2}} = 0,
\end{equation}
which leads to
\begin{equation}
    \frac{\omega'}{c} = \pm \left(k_x' + \frac{2\pi n}{p'} \right),
\end{equation}
where $p' = \gamma p$ is the dilated grid period in the co-moving frame. In order to find the corresponding diffraction resonances in the laboratory frame, we use the relativistic Doppler transformation formulas \eqref{lorentz_omega} and \eqref{lorentz_kx}. Considering normal incidence ($k_x = 0$), the diffraction resonances will be located at the (angular) frequencies
\begin{equation} \label{resonance_normal}
    \omega_\textrm{c} = \frac{2\pi nc}{\gamma^2 p \left(1 \pm \frac{v}{c} \right)}.
\end{equation}
Without the synthetic motion ($v=0$, $\gamma =1$), the above equation reduces to the well-known expression for static gratings ($\omega_c = \pm 2\pi n c / p)$. The spacetime modulation ($v>0$) of the metallic grating originates a Doppler shift of the diffraction resonances, which is mostly relevant if the modulation speed becomes a significant fraction of the speed of light. In particular, the synthetic motion causes the \textit{splitting} of the original diffraction resonances into two different resonances (see the term $1 \pm v/c$). For small modulation speeds, the two resonances are nearly equally-spaced with respect to the original static resonance. 

\subsection{Electromagnetic fields diffracted by the spacetime grating}

In the following, we use the Lorentz transformation to characterize some features of the exact fields (beyond the homogenization limit) diffracted by the spacetime grating. 

\subsubsection{Field profiles}
It was shown in previous works that the excited tangential electric field profile in a static metallic grating is roughly proportional to \cite{FloquetCircuit1, FloquetCircuit2}
\begin{equation}
f_t(x) \propto \begin{cases} \left[1 - \left(\frac{2x}{w}\right)^2 \right]^{-1/2} , & \text {TM pol.} \vspace*{0.2cm} \\
\left[1 - \left(\frac{2x}{w}\right)^2 \right]^{1/2}, & \text {TE pol.} \end{cases}, 
\end{equation}
over the air slit width ($-w/2 \leq x \leq w/2$) and zero elsewhere, in a unit cell of spatial period $p$. We can take advantage of the properties of the Dirac impulse train of period $p$, $D(x) = \sum_{k=-\infty}^\infty \delta(x - kp) $, and of the convolution operator (denoted by the symbol $*$) to extend the electric field profile from a single unit cell to the entire grid, as follows:
\begin{equation}
    E_t(x) \propto  D(x) * f_t(x) , \quad -\infty \leq x \leq \infty
\end{equation}

For a moving grating with uniform speed $v$, the electric field profile $E_t$ suffers a translation by $v t$  along the direction of motion  as time advances. This can be mathematically expressed as

\begin{equation} \label{Et_xt}
E_t(x, t) \propto D(x-vt) * f_t(x),
\end{equation}
with time period $T = p/v$. Therefore, to a first approximation, the moving grating may be replaced by an infinitesimally-thin screen whose field profile is described in Eq.\eqref{Et_xt}. This result can serve as a basis to analyze the moving grating via Floquet-Bloch expansions (taking advantadge of the spacetime periodicity) together with integral-equation methods \cite{Timevarying2022, Timevarying2023}.

\subsubsection{Floquet-Bloch expansions and spectral relations}
The Lorentz transformation can be used to establish a simple link between the Floquet-Bloch expansions of the electromagnetic fields in the laboratory and co-moving frames. In the laboratory frame, the Floquet-Bloch expansion of the transverse electric field can be represented as follows \cite{Taravati2019}:
\begin{equation}\label{expansion_lab}
E_x= \displaystyle\sum_{\forall n}\displaystyle\sum_{\forall m} E_{nm} \text{e}^{-\text{j}( k_{x m} x - \omega_{n}t)}   \hspace{5 mm} n, m \in \mathbb Z 
\end{equation}
where $k_{xm} = k_x + \dfrac{2\pi m}{p}$ and $\omega_{n} = \omega + \dfrac{2 n \pi}{T} $ determine the space and time dependence of each harmonic. In the co-moving frame, since the grating is at ``rest'', only space harmonics take part on the field expansion:
\begin{equation}\label{expansion_com}
E_x^{'} = \displaystyle\sum_{\forall q} E'_{q} \, \text{e}^{-\text{j} (k'_{xq} x' - \omega' t')} \hspace{5 mm} q \in \mathbb Z   
\end{equation}
with $k_{xq}' = k'_x + \dfrac{2  \pi q}{p'}$. Taking into account the invariance of the transverse fields along the direction of motion, 
\begin{equation}
E_x = E'_x \, ,
\end{equation}
 it is straightforward to establish the following relation between the harmonics in the laboratory and co-moving frames:
\begin{align}
E_{nn} &= E_{q}' \hspace{5 mm} \text{only} \hspace{1mm} \text{if} \hspace{5 mm} n = q \\ 
E_{nm} &= 0 \hspace{7 mm} \text{other}
\end{align}

Thus, \eqref{expansion_lab} can be rewritten as
\begin{equation}\label{expansion_lab2}
E_x = \displaystyle\sum_{\forall n} E_{n}\, \text{e}^{-\text{\j} (k_{xn} x - \omega_{n}t)}    \,,
\end{equation}
where the notation $E_{nn}$ (case $m=n$) for Floquet harmonics has been simplified to $E_n$. Thereby, the invariance of the electric field component parallel to the direction of motion ($E_x = E_x'$) implies that the corresponding Floquet harmonics have identical coefficients in the laboratory and co-moving frames $E_{qq} = E_{q}'$. In particular, the phase invariance of the fields $E_x, E_x'$ implies the phase invariance of the coefficients of the individual harmonics in different frames.  

This result allows for a simple calculation of the cutoff frequency associated with harmonics in the laboratory frame. Defining the propagation constant (perpendicular to the grating) of a given harmonic as
\begin{equation}
    k_{y n} = \sqrt{(\omega_{n}/c)^{2} - k_{xn}^{2}}\,,
\end{equation}
the cutoff frequency (c) of each harmonic is obtained when $k_{\text{y}n} = 0$, leading to the condition:
\begin{equation}
\omega_\textrm{c} = \pm \frac{2n\pi c}{p} \frac{1 \mp v/c}{ 1 \mp \sin(\theta)}   
\end{equation}
For normal incidence, the former expression reduces to
\begin{align}
  \omega_\textrm{c} &= \frac{2n\pi c}{p} (1 - v/c) = \frac{2n\pi c}{p} \frac{1}{\gamma^{2}(1 + v/c)}, \hspace{4mm} n > 0 \\  
\omega_\textrm{c} &= \frac{-2n\pi c}{p} (1 + v/c) = \frac{2n\pi c}{p} \frac{1}{\gamma^{2}(1 - v/c)}, \hspace{1mm} n < 0
\end{align}
These expressions agree with Eq. \eqref{resonance_normal}, giving consistency to the equivalent circuit development and its implications. 

\section{Numerical Examples}
This section presents a numerical study intended to validate and illustrate the application of the developed theory. Specifically, we calculate the reflection and transmission coefficients when the spacetime grating is illuminated by an incoming wave, considering both normal and oblique incidences. Furthermore, we will extend our theory to more complex scenarios, showing that a moving grating backed by a metallic plate can be also studied with similar analytical methods.

\subsection{Normal incidence}

\begin{figure}[!t]
	\centering
	\subfigure[]{
		\includegraphics[width=0.5\textwidth]{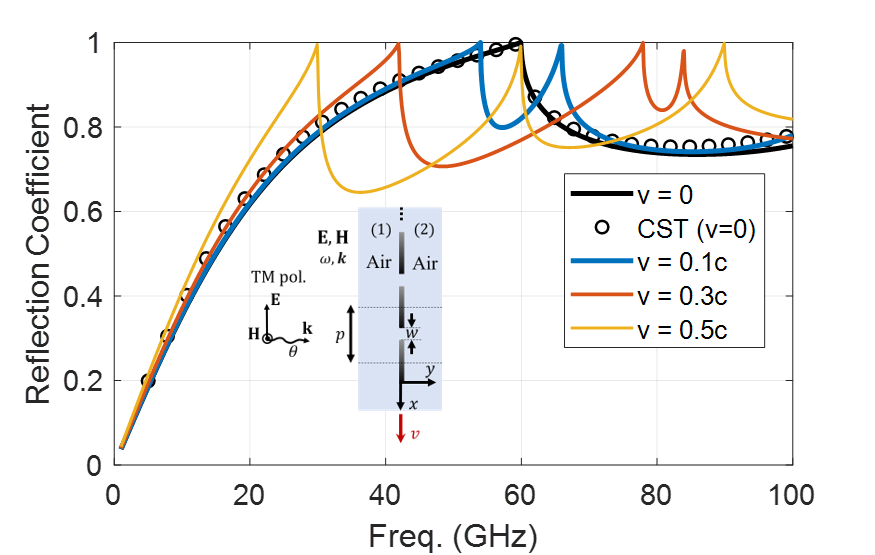}}\\
    \vspace{-0.5cm}\subfigure[]{
		\includegraphics[width=0.5\textwidth]{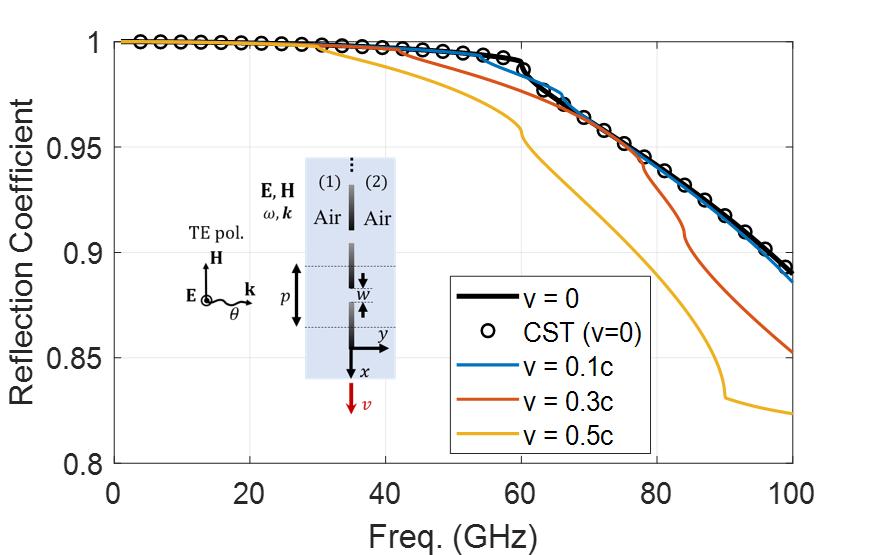}}\\
	\caption{\small Magnitude of the reflection coefficient as a function of the modulation speed $v$ of the metallic grating. (a) TM normal incidence. (b) TE normal incidence. For the static case ($v=0$), a comparison is provided with the commercial software CST.  }
	\label{fig2}
\end{figure}

The inset of Figure~\ref{fig2}(a) illustrates the geometry of the first example. A TM-polarized plane wave illuminates the moving grating. The wave propagates along $+y$ so that the incidence is along the normal direction ($\theta = 0$). The period of the grating is $p = 5$ mm and the slit width is $w = 1$ mm. The grating is periodic along the $x$-axis, which is also the direction of the synthetic motion. The reflection coefficient $R$ is calculated using \eqref{R} for different velocities $v$. For the static scenario ($v=0$), the analytical results are compared with the commercial simulator CST Microwave Studio, which serves as a benchmark for accuracy. An excellent agreement is observed between the analytical approach and CST, even far beyond the onset of the diffraction regime (60 GHz).

When the spacetime modulation is switched on, the interactions between the wave and the grating are effectively mediated by a Doppler-like (frequency-shift) effect. Indeed,  the diffraction resonance, originally located at 60 GHz in the static case is split into two resonances shifted upwards and downwards in frequency due to the synthetic motion [see Figure \ref{fig2}(a)].  The higher is the considered modulation speed, the farther the resonance moves away. Actually, the diffraction resonance "splitting" can be explained qualitatively as follows.  A normal incidence in the laboratory frame ($\theta = 0$) does not imply normal incidence in the co-moving frame ($\theta' \neq 0$). 
In fact, using Eq.\eqref{lorentz_kx} it can be easily demonstrated that the incidence angle $\theta'$ in the co-moving frame is
\begin{equation} \label{theta_prime}
    \theta' = 
    \arcsin \left( \frac{ {\sin \theta  - \frac{v}{c}}}{1-\frac{v}{c}\sin \theta} \right) \underset{v,\theta \ll 1}{\approx } 
    \arcsin \left( {\sin \theta  - \frac{v}{c}} \right),
\end{equation}
which is evidently nonzero when $\theta=0$. The diffraction resonance splitting can be attributed to the nonzero $\theta'$, which leads to diffraction resonances of order $n=\pm 1$ with a different frequency. 

As discussed previously, the frequency of the diffraction resonances can be calculated analytically by using \eqref{resonance_normal}. The theoretical values are $f_{1} = 60$ GHz ($v=0$), $f_{1} = 66$ GHz ($v=0.1c$), $f_{1} = 78$ GHz ($v=0.3c$) and $f_{1} = 90$ GHz ($v=0.5c$) for the $n=1$ harmonics. For the diffraction order $n=-1$, resonances appear at $f_{-1} = 60$ GHz ($v=0$), $f_{-1} = 54$ GHz ($v=0.1c$), $f_{-1} = 42$ GHz ($v=0.3c$) and $f_{-1} = 30$ GHz ($v=0.5c$). These theoretical calculations are in excellent agreement with the numerical results plotted in Figure \ref{fig2}(a). In addition, there are two diffraction resonances located at $f_{2} = 60$ GHz ($v=0.5c$, yellow curve) and $f_{2} = 84$ GHz ($v=0.3c$, red curve), which are to the second-order spatial harmonic $n=2$. Thus, the numerical analysis confirms that for normal incidence the diffraction resonances are symmetrically shifted in frequency.  It is worth underlining that the currently available commercial solvers are unable to analyze moving and time-modulated gratings. This underscores the novelty and significance of the approach introduced in our work.

Figure \ref{fig2}(b) shows the magnitude of the reflection coefficient when the spacetime grating is illuminated by a normal incident TE-polarized  plane wave.  In case of TE incidence, the diffraction resonances coincide with those in the TM case. In the TE case, the spacetime grating shows a mainly reflective behavior, as opposed to the mainly transmissive behavior of the TM case at low frequencies. 

Although not explicitly shown in the figures, we have checked that the energy conservation condition \linebreak ($|R|^2 + |T|^2 = 1$ below the diffraction regime) is satisfied in the homogenization limit. The energy-conservation results from the continuity of the Poynting vector through the slit in the laboratory frame. The continuity is satisfied due to the negligible thickness of the grating. Note also that due to the PEC boundary condition the normal component of the Poynting vector vanishes at the moving metallic grating. For thick gratings, the continuity of the Poynting vector across the interface cannot be assumed, and in that case the response may become active. Above the onset of diffraction, the power transported by higher-order modes needs to be taken into account in the energy balance, as higher-order modes are no longer evanescent in that case.

\subsection{Oblique incidence. Nonreciprocity}
\begin{figure}[!t]
	\centering
	\subfigure[]{
		\includegraphics[width=0.45\textwidth]{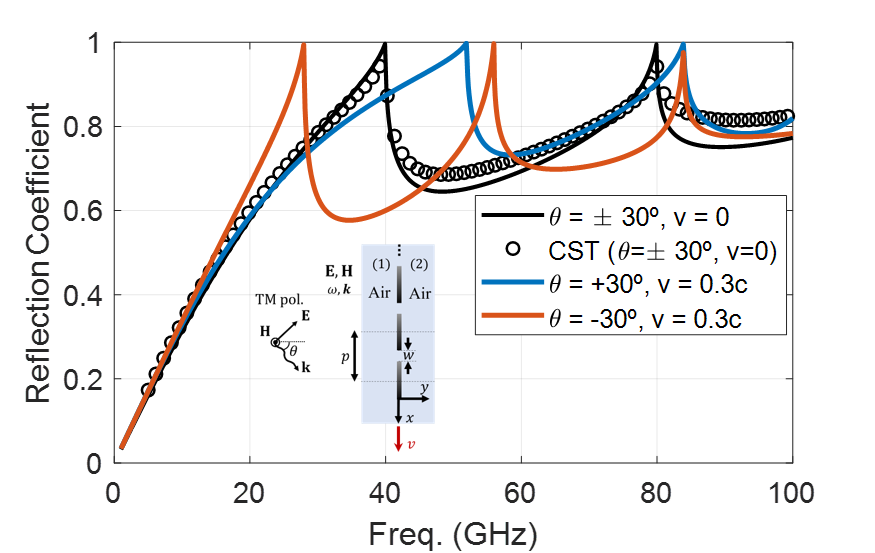}}\\ \vspace{-0.5cm}
	\subfigure[]{\vspace{-0.3cm}
		\includegraphics[width=0.45\textwidth]{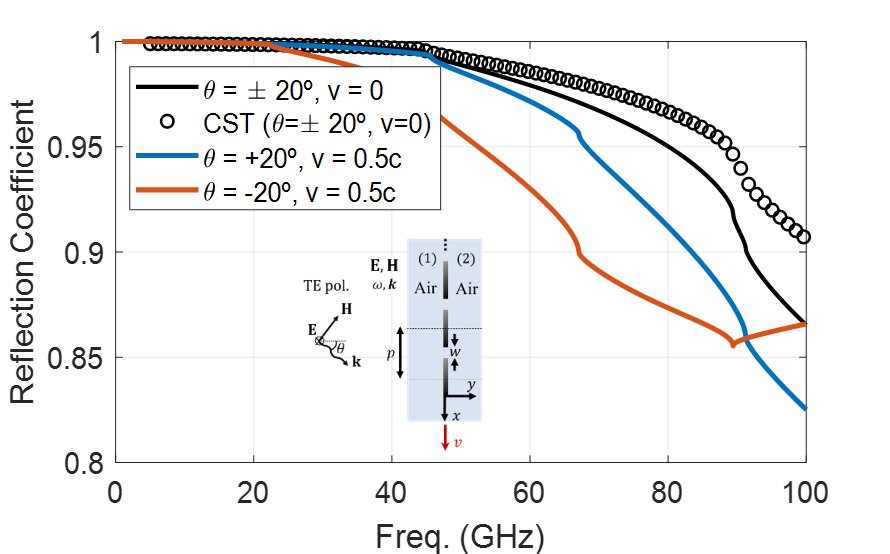}}\\	
	\caption{\small Magnitude of the reflection coefficient as a function of the modulation speed $v$ of the metallic grating. (a) TM oblique incidence ($\theta = \pm 30^\mathrm{o}$). (b) TE oblique incidence ($\theta = \pm 20^\mathrm{o}$). For the static case ($v=0$), a comparison is provided with the commercial software CST Microwave Studio. } 
	\label{fig3}
\end{figure}

Figure \ref{fig3}(a) considers the case where a TM-polarized wave obliquely ($\theta = \pm 30^\mathrm{o}$) impinges on the moving grating. The dimensions of the grating remain as in the previous example ($p = 5$ mm, $w=1$ mm). Similar to the previous case, for the static case ($v= 0$) there is a good agreement between CST and our numerical approach over a broad bandwidth, even beyond the diffraction regime (40 GHz when $\theta = 30^\mathrm{o}$). Note that when $v=0$ the system is reciprocal, and consequently $R(\omega, \theta) = R(\omega, -\theta) $. However, for a moving grating the electromagnetic reciprocity is broken, and thereby it is possible that $R(\omega, \theta) \neq R(\omega, -\theta) $. This scattering asymmetry is quite evident in the numerical simulations for the cases $\theta = + 30^\mathrm{o}$ (blue curve) and  $\theta = - 30^\mathrm{o}$ (red curve) when $v=0.3c$, which are no longer coincident. 

Formally, the nonreciprocity of the system can be explained through time reversal symmetry breaking originated by the synthetic motion \cite{Caloz2018, TimeSymmetry2014}. In fact, under a time reversal the modulation speed of the grating changes sign. Thus, the time modulation leads to  a situation where the effective response of the grating depends on the sign of the direction of arrival $\pm \theta$ of the incoming wave. 

Figure \ref{fig3}(b) considers the case where a TE-polarized wave obliquely ($\theta = \pm 20^\mathrm{o}$) impinges on the moving grating. As before, there is a good agreement with full wave simulations in the static case. Nonetheless, slight differences appear above the first grating-lobe cutoff (60 GHz).
Moreover, also in this case the nonreciprocal response is observed when the spacetime modulation is switched on. Evidently, the reflection coefficient is expected to differ significantly for the TM and TE polarization illuminations, as in the latter case the incident electric field is parallel to the metallic strips, whereas in the former case it is not. Consistent with this property, Figs. \ref{fig3}(a) and \ref{fig3}(b) show that the grid scatters weakly (strongly) for TM (TE) polarized incident waves, respectively, at low frequencies, independent of the modulation speed. 

\subsection{Galilean vs Lorentz Doppler transformations}

Galilean transformations (see Appendix C for further information) may also be used to switch to a co-moving frame where the material response becomes time independent \cite{Galiffi2022, paloma_spacetime}. 
However, the free-space Maxwell equations are not invariant under a Galilean transformation. In particular, under a Galilean transformation the air regions would be transformed into a material with a bianisotropic response \cite{Galiffi2022, paloma_spacetime}, which evidently greatly complicates the electromagnetic analysis in the co-moving frame. Thus, the Galilean transformation leads to a cumbersome formalism in the scenarios studied in this work, and hence will not be considered here in detail. 

It is however instructive to see the impact of Galilean Doppler transfomation (see Appendix C) in Eqs. \eqref{hom5} and \eqref{homTE} as opposed to the relativistic (Lorentz) Doppler-transformation considered thus far. The Galilean Doppler shift is based on the formulas $\omega' = \omega - v k_x$ and $k'_x = k_x$.

\begin{figure}[!t]
	\centering
	\subfigure[\hspace{-0.6cm}]{
		\includegraphics[width=0.23\textwidth]{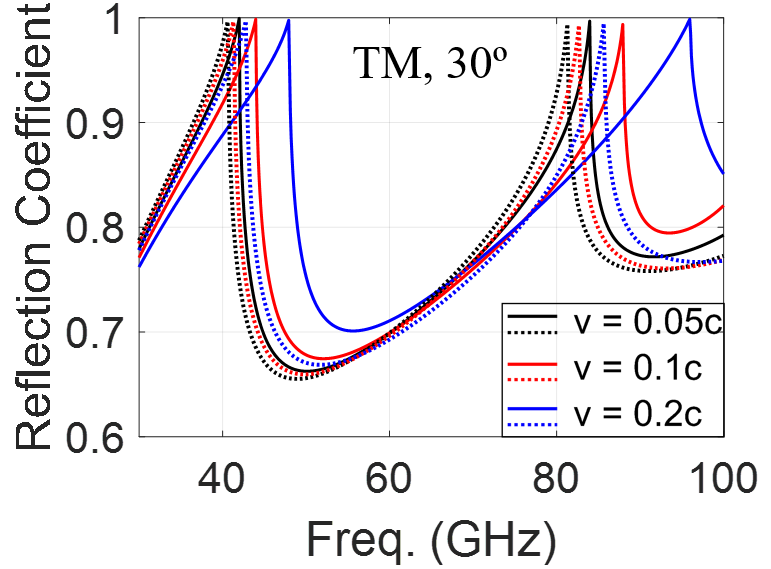}}
	\subfigure[\hspace{-0.6cm}]{
		\includegraphics[width=0.23\textwidth]{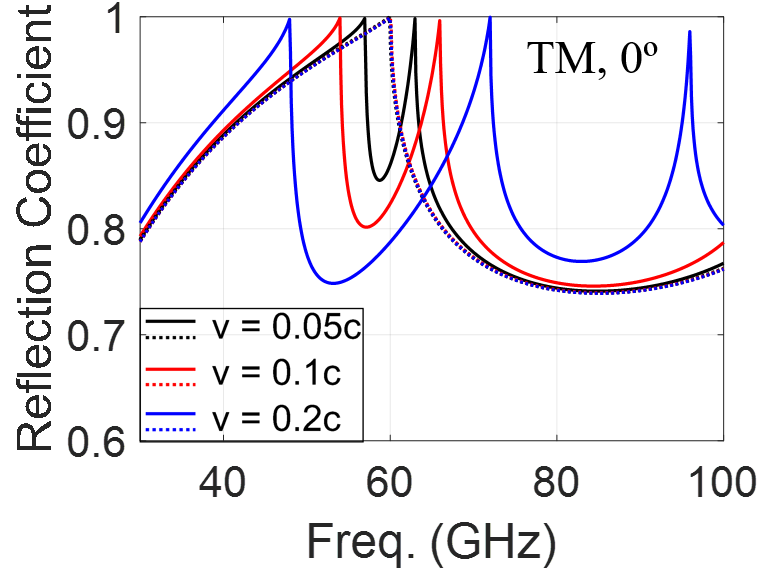}}\\	
	\caption{\small Numerical comparison between the reflection coefficient calculated using the Galilean Doppler transformation (dashed lines) and the exact relativistic (Lorentz) Doppler theory (solid lines). (a) TM oblique incidence ($\theta = 30^\mathrm{o}$). (b) TM normal incidence ($\theta = 0^\mathrm{o}$). In panel (b), all the dashed curves are coincident. Grating parameters: $p = 5$ mm, $w = 1$ mm. } 
	\label{galilean}
\end{figure}

Figure \ref{galilean} shows a numerical comparison between the reflection coefficient calculated  using either the Galilean or Lorentz Doppler-shifts in the grid equivalent admittance.  For the case of oblique TM incidence, it can be seen in Figure \ref{galilean}(a) that the Galilean results start to deviate significantly from the  rigorous Lorentz approach for velocities larger than $0.1c$. A similar trend was observed for other metallic gratings with different dimensions.

On the other hand,  Figure \ref{galilean}(b) shows that for normal TM incidence, the Galilean approximation fails to capture the splitting of the diffraction resonances around 60 GHz, different from the rigorous relativistic theory. This can be easily explained noting that under normal incidence conditions ($k_x=0$), the Galilean Doppler transformation predicts erroneously that $\omega' = \omega$ and $k_x' = k_x$. Consequently, the results obtained with this approximation are completely insensitive to the modulation speed for normal incidence, very different from the rigorous relativistic treatment of the problem via the Lorentz transformations.


\subsection{Spacetime-modulated gratings backed by a metallic plate}
\begin{figure}[!t]
	\centering
	\subfigure{ \hspace{-0.8cm}
		\includegraphics[width=0.54\textwidth]{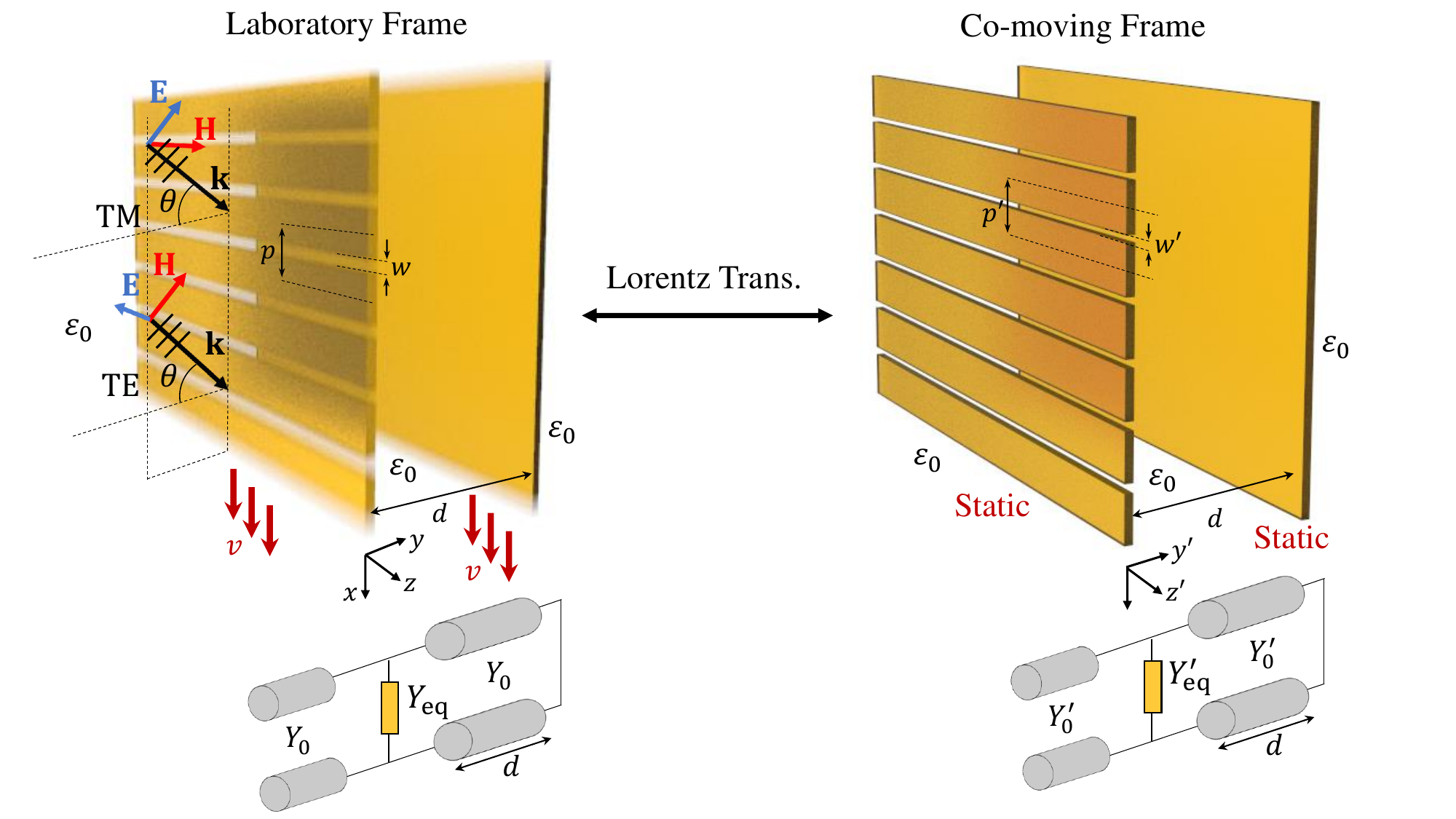}} \vspace{-0.5cm}\\
	\caption{\small Moving metal grating backed by a metallic plate. The original (laboratory) and transformed (co-moving) frames are shown.  Their parameters are related via a Lorentz transformation. } 
	\label{fig4}
\end{figure}

Next, we consider the scenario where a moving (spacetime modulated) grating is backed by a metallic (PEC) plate, as sketched in Figure \ref{fig4}. Some space-modulated (time-invariant) cases have been already treated in the literature \cite{FloquetCircuit3, Molero_meander2017}. The distance between the grating and the metal plate is $d$. Since the materials that form this system (air and PEC thin plates) have a Lorentz invariant response, the response in the co-moving frame is the same as in the static case, apart from the length dilation discussed previously. In this problem, the system acts as a fully-reflective structure, since the metal loss is neglected. 

It is straightforward to derive the equivalent circuit model of this system.  Evidently, the air semi-space on the left is associated with the input transmission line (analogous the free-standing grating), the moving grid is modelled by the admittance $Y_{\rm{eq}}$, whereas the metal plate is modelled by a short-circuit. It should be noted that for a small $d$ the back metallic plate may affect the value of the equivalent admittance compared to the free-standing case. This may happen due to the interaction of higher-order evanescent diffraction modes with the back plate; 
In contrast, when the separation $d$ is electrically large, the equivalent admittance remains unaltered as compared to the free-standing case. See Appendix D for further details. 

\subsubsection{Goos-H\"anchen shift}

\begin{figure}[!t]
	\centering
	\subfigure[]{
		\includegraphics[width=0.52\textwidth]{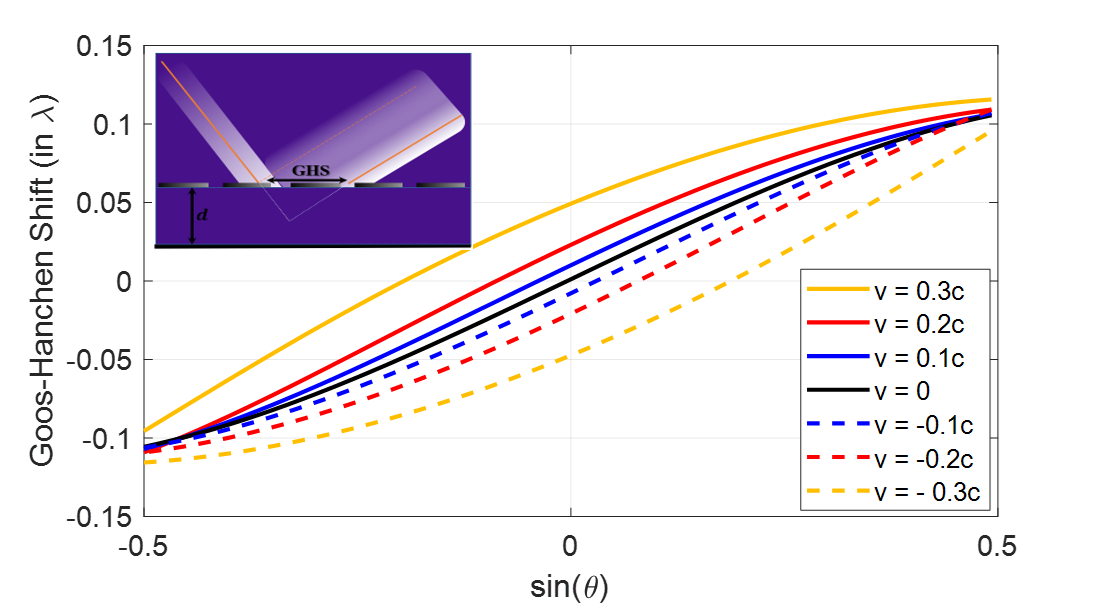}}\\
  \vspace{-0.5cm}\subfigure[]{
		\includegraphics[width=0.52\textwidth]{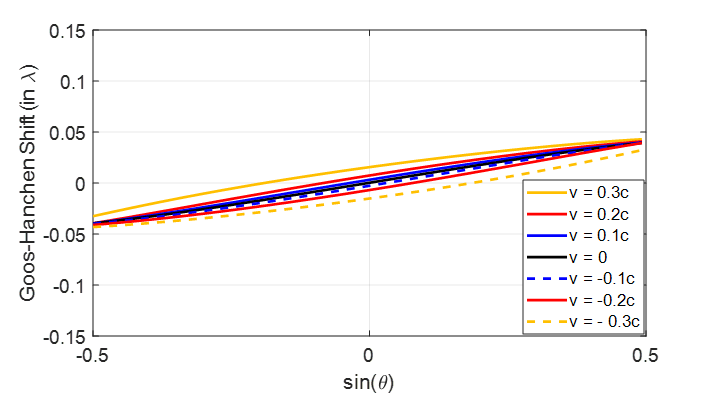}}\\
	\caption{\small Goos-Hänchen shift (in electrical units) suffered by a wave illuminating  a moving metallic grating backed by a metallic plate, as a function of the modulation speed $v$ for several incidence angles $\theta$.  The grating and the metallic plate are separated a distance of (a) $d = 0.1p$ and (b) $d = 0.05p$. TM incidence is assumed. The operation frequency is 20 GHz.}
	\label{fig5}
\end{figure}

A light beam impinging on a material interface or on a fully-reflective metasurface is prone to suffer the so-called Goos-Hänchen shift (GHS) \cite{GHshift}. The \emph{spatial} Goos-Hänchen (GH) shift
refers to the transverse displacement of a reflected beam along the interface ($x$ axis) of a structure with respect to the point of reflection that would be expected in a perfectly specular reflection.
During the last decades, this phenomenon has attracted the attention of scientists and engineers in plasmonics, metamaterials, and two-dimensional materials such as graphene and others \cite{GHintro1, Lakhtakia2004, GHintro3}. The inset of Figure \ref{fig5} illustrates a \emph{positive} GH shift. However, \emph{negative} GH shifts are also possible without violation of causality \cite{GH_negative1, GH_negative2}.

The Goos-Hänchen shift can be derived from the phase $\varphi$ of the reflection coefficient, $R = |R|\, \mathrm{e}^{j \varphi}$, \cite{GHshift1}
\begin{equation}
    \mathrm{GHS} = - \frac{\partial \varphi}{\partial k_x} =  - \frac{1}{k_0} \, \cdot \frac{\partial \varphi }{\partial \sin\theta}.
\end{equation}
Heuristically, one may expect that the GH shift created by a spacetime modulated screen should be tailored by the modulation speed, as the interaction of waves with ``moving'' structures typically leads to a (synthetic) Fresnel drag of the radiation towards the direction of motion \cite{Huidobro2019}.

In order to investigate this effect, we characterized the GH shift (${\Delta _{{\rm{GH}}}}$) in a system formed by the moving grating backed by a metallic screen (Figure \ref{fig5}). As previously mentioned, for this geometry the amplitude of the reflection coefficient is exactly one ($|R|=1$). 
As seen in Figure \ref{fig5}, when oblique incidence is considered ($\theta \neq 0$), the incident beam is shifted parallel to the interface where the moving grating is placed.

When the grating is static, the GH-shift (black curve) is perfectly symmetric for positive and negative incident angles: ${\Delta _{{\rm{GH}}}}\left( \theta  \right) =  - {\Delta _{{\rm{GH}}}}\left( { - \theta } \right)$. On the other hand, when the spacetime modulation is switched on ($v\neq 0$),  the odd symmetry of the GH-shift is broken, evidencing the nonreciprocal response of the system. In the time-variant case, the following symmetry ${\Delta _{{\rm{GH}}}}\left( {\theta ;v} \right) =  - {\Delta _{{\rm{GH}}}}\left( { - \theta ; - v} \right)$ is observed. Thereby, negative velocities (the grating ``moves'' along the $-x$ axis) change the trend of the GH shift compared to positive velocities.

Remarkably, even for normal incidence ($\theta=0$), the spatial GH shift does not vanish when the grating is moving. This can be understood noting that a normal incidence in the laboratory frame corresponds to an oblique incidence in the co-moving frame, as seen in Eq.\eqref{theta_prime} (see Sec. III.A). Normal incidence GH shift has been discussed in previous works on plasmonics and gradient metasurfaces \cite{GH_normal1, GH_normal2}. From a different point of view, the nontrivial GH shift may be regarded as a consequence of the synthetic Fresnel drag provided by the moving grid. In fact, as can be seen in Figures \ref{fig5}(a) and \ref{fig5}(b), positive values of $v$ tend to favour positive shifts ${\Delta _{{\rm{GH}}}}$ due to the synthetic motion of the grating. For example for $v=0.3c$ one sees that the GH-shift is positive for incidence angles larger than $\theta = -11.5^\mathrm{o}$ in Figure \ref{fig5}(a).

Figure \ref{fig5}(b) illustrates the GH shift when the  spacetime grating and the back metallic plate are a little closer than in Figure \ref{fig5}(a). When the separation $d$ between grating and back plate is very small, the phase $\varphi$ of the reflection coefficient approaches the constant value of 180 deg. Therefore, the GH shift created by the spacetime system progressively diminishes as the value of $d$ decreases. Note that in the extreme case $d=0$, full reflection and a constant phase of 180 deg would be observed, since the back metallic plate would overlap with the grating, leading to a null GH shift. 

From a practical perspective, synthetic motion can be added to the grid in the form of travelling-wave spacetime modulation schemes \cite{Mazor2020, Tretyakov2021}. For this goal, the metal strips that form the grid can be loaded with reconfigurable lumped elements, such as time-modulated capacitors or varactor diodes \cite{Kreiczer2021, Wu2020, Jayathurathnage2021}. This would be an appropriate solution for radiofrequency (RF) designs. In the THz regime, \textit{p-n} junction diodes or transistor-compatible devices may be utilized \cite{Lira2012, Mikheeva2022}.

\section{Conclusion}

In this paper, we have presented an analytical framework for the study of moving or spacetime-modulated metallic gratings. Lorentz transformations allow us to analyze the system in a frame co-moving with the material (or with the spacetime modulation), where the metallic grating has a static-like response. In the co-moving frame, the system is time-invariant and thereby it is possible to apply well-established circuit-theory methods to characterize the response of the static (modulated only in space) grating. The effective response in the original laboratory frame can then be determined using a Lorentz transformation. Thus, our formalism provides a simple but rather powerful tool to characterize the scattering parameters and dynamical response of moving or spacetime-modulated metallic gratings. 

We have derived analytical expressions for the diffraction resonances created by the moving grating. We have numerically observed the Doppler-like shift and the splitting of the diffraction resonances that the synthetic motion originates. Additionally, we highlighted that the nonreciprocity of the system leads to a strongly asymmetric scattering response, such that the reflection levels for the $\pm \theta$ angles may differ dramatically. 
Subsequently, we have shown that for velocities larger than $0.1c$ it is necessary to fully take into account the relativistic corrections in the Doppler effect, as a Galilean approximation leads to erroneous results. Finally, we have analyzed the case of a moving grating backed by a metal plate. This fully-reflective moving system leads to spatial Goos-Hänchen shift tailored by the velocity of the grid. We have shown that the reflected beam tends to be dragged towards the direction of the synthetic motion, leading to a Goos-Hänchen shift even under normal incidence conditions. Furthermore, the synthetic motion can also enable negative GH shifts, such that the sign of the GH-shift is different from the sign of the incident angle $\theta$. 

We believe that our findings shed new light on the properties of moving and spacetime-modulated metallic gratings and are anticipated to contribute to a better understanding of the physical phenomena underlying these complex structures. Additionally,  novel mathematical tools developed herein are expected to be of potential interest to the engineering and physics communities.

\begin{acknowledgments}
A. A.-A. and C. M. were supported in part by the Spanish Government under project PID2020-112545RB-C54 and project RTI2018-102002-A-I00, in part by ``Junta de Andalucía'' under project B-TIC-402-UGR18, project A-TIC-608-UGR20, project PYC20-RE-012-UGR and project P18.RT.4830. M. G. S. was supported in part by the Institution of Engineering and Technology (IET), by
the Simons Foundation, and by Fundação para a Ciência e a Tecnologia and Instituto de Telecomunicações
under project UIDB/50008/2020.

\end{acknowledgments}

\appendix

\section{Lorentz-invariance of PEC material}
Let us consider an infinitesimally-thin PEC plate lying on the XOZ plane. Thus, the  tangential electric fields  must vanish at the boundaries in the laboratory (unprimed) frame, as well as the normal component of the magnetic field:
\begin{equation}
    E_x = E_z =  0,\quad \quad B_y = 0\, .
\end{equation}
Equation \eqref{fields1} directly dictates that the $x$ component in the co-moving (primed) frame must vanish for the considered PEC case:
\begin{equation}
     E_x' = E_x = 0\, .
\end{equation}
Moreover, by applying Eq. \eqref{fields3} to the present case, with $E_z = 0$ and $B_y = 0$, we find that $E'_z$ also vanishes:
\begin{equation}
    E_z' = \gamma \Big(E_z + vB_y\Big) = 0\, .
\end{equation}
Thus, the tangential electric fields vanish for a PEC plate in both laboratory and co-moving frames. Therefore, PEC regions are frame-invariant materials. The same conclusions would be obtained by operating in the reverse direction, that is, by fixing $E'_x = E'_z = 0$ and then retrieving the fields $E_x = E_z = 0$ in the laboratory frame. Note that the former rationale holds as long as the considered spacetime grating is infinitesimally thin (in practice, electrically thin) along the $y$ direction. PEC regions in thick spacetime gratings cannot be considered to be frame-invariant. In fact, when the normal to a PEC plate (${\bf{\hat n}}$) is along the direction of motion, it can be shown that the PEC boundary condition in the laboratory frame (${\bf{\hat n}} \times {\bf{E}} = 0$), results in a boundary condition of the type ${\bf{\hat n}} \times \left( {{\bf{E'}} - {\bf{v}} \times {\bf{B'}}} \right) = 0$ in the co-moving frame.

\section{Equivalent admittance in the co-moving frame ($Y'_\mathrm{eq}$)} \label{sec:appendixa}

Following the discussion presented in Section \ref{sec:analytical}, the moving metallic grating can be represented in the laboratory frame as an equivalent admittance $Y_\mathrm{eq}$, which is determined by the equivalent admittance $Y'_\mathrm{eq}$ in the co-moving frame (static grating). The theory to analyze \emph{static} metallic metagratings was developed for 1-D periodic structures in \cite{ FloquetCircuit1, FloquetCircuit2, FloquetCircuit3}, for 2-D structures in \cite{FloquetCircuit0, FloquetCircuit_2D, FloquetCircuit_2D1, FloquetCircuit_2D2} and, recently, for 3-D structures \cite{Alex3D_2023}. According to \cite{FloquetCircuit1}, the equivalent admittance of a 1-D grating is given by:
\begin{equation} \label{Yeq_prime}
    Y'_\mathrm{eq}= \hspace*{-0.3cm}
    \sum_{\footnotesize \begin{array}{cc} n=-\infty \\ {n \neq 0} \end{array}}^\infty \hspace*{-0.3cm}
    N_n^{'2} \left( Y_{n}^{'(1)}+Y_{n}^{'(2)} \right) 
\end{equation}
where 
\begin{equation} \label{Yn_prime}
Y_{n}^{'(i)}= \begin{cases}\frac{\omega' \varepsilon_0 \varepsilon_r^{(i)}}{ k_{yn}^{'(i)}}, & \text {TM pol.} \vspace*{0.2cm} \\
\frac{ k_{yn}^{'(i)}}{\omega' \mu_{0}}, & \text {TE pol.} \end{cases}
\end{equation}
is the wave admittance of the input ($i=1)$ and output ($i=2$) media for the two incident polarizations (TM and TE), represented here as semi-infinite transmission lines.  For the case of a free-standing grating, the input and output regions are air. In such a case, $Y_{n}^{'(1)} = Y_{n}^{'(2)}$, $\varepsilon_r^{(i)} = 1$, which allows for some simplification of \eqref{Yeq_prime}. It is worth to remark that expression \eqref{Yeq_prime} is known to work well even beyond the onset of the diffraction regime and for any incident angle $\theta$. 

In \eqref{Yn_prime}, $k_{yn}^{'(i)}$ is the propagation constant of the air regions that can be computed as
\begin{equation}
    k_{yn}^{'(i)} = \sqrt{\varepsilon_r^{(i)} k_0^{'2} - k_{xn}^{'2}}
\end{equation}
where
\begin{equation}
    k'_{xn} = k'_x + k'_n 
\end{equation}
is the transverse wavenumber associated with the $n$-th spatial harmonic. $k'_x$ can be easily obtained from $k_x = k_0 \sin(\theta)$ by applying the relativistic Doppler  transformation, while $k'_n$ is given by $k'_n = 2 \pi n / p'$ with $p' = \gamma p$ being the period of the static metallic grating.

The term $N'_n$ in \eqref{Yeq_prime} is defined as
\begin{equation}
    N'_n= \begin{cases}\frac{J_{0}\left(k'_{xn} w' / 2\right)}{J_{0}\left(k'_{x} w' / 2\right)} & \text {TM} \vspace*{0.2cm} \\  \frac{J_{1}\left(k'_{xn} w' / 2\right)}{J_{1}\left(k'_{x} w' / 2\right)} \cdot \frac{k'_{x}}{k'_{xn}} & \text {TE}\end{cases}
\end{equation}
where $J_0(\cdot)$ and $J_1(\cdot)$ are the Bessel functions of the first kind of order zero and one, respectively, and $w' = \gamma w$ is the primed slit width. The circuit model is known to give accurate results as long  as $w' \lesssim 0.6p'$. $N'_n$ can be seen as a transformer ratio associated with higher-order Floquet harmonics that relates the spatial spectrum of
the fields at the slit region \cite{FloquetCircuit1}. Note that due to the translation invariance of the problem along the direction of the metal strips ($z$-direction), TM/TE incident waves can only excite TM/TE harmonics at the discontinuity plane ($y$-direction), respectively.

\section{Galilean Doppler-transformation}
Similarly to Lorentz transformations, it is possible to use a Galilean transformation to analyze the moving grating depicted in Fig. \ref{fig1}.  Assuming that the grating moves with uniform velocity $v$ along the $x$ direction, the Galilean transformation links the laboratory frame (unprimed) coordinates to the co-moving frame (primed) coordinates as follows:
\begin{equation}
    x' = x - vt, \quad y' = y, \quad z' = z, \quad t' = t,
\end{equation}
The Galilean transformation assumes that time is absolute, meaning that time intervals are the same for all observers, independent of the relative motion. The Galilean spectral transformations are expressed as (Galilean-Doppler transformation)
\begin{equation}
    \omega ' = \omega - vk_x, \quad \mathbf{k}' = \mathbf{k}.
\end{equation}
Thus, $k'_x = k_x$; In particular, the wavelength and the direction of the wave vector are preserved by the Galilean transformation. 

As is well known, the free-space Maxwell equations are not invariant under Galilean transformations. 
This property undermines the usefulness of the Galilean transformation in the context of our work, as it implies that air (vacuum) is not transformed into itself under a Galilean transformation, but rather into a bianisotropic material.


\section{Equivalent admittance ($Y'_\mathrm{eq}$) and reflection coefficient ($R$) in a metal-backed metallic grid} \label{sec:appendixB}

Let us consider that the output region ($i=2$) is backed by a metal plate. Thus, the wave admittance $Y_{n}^{'(2)}$ in Eq.\eqref{Yeq_prime} should take into account the effect of the metal plate. This can be easily accounted for by changing the value of $Y_{n}^{'(2)}$ in Eq.\eqref{Yeq_prime} by the term $Y_{n}^{'\mathrm{(R)}}$, defined as
\begin{equation} \label{Ynprime_short}
    Y_{n}^{'(2)} \rightarrow Y_{n}^{'\mathrm{(R)}} = -j Y_{n}^{'(2)} \cot \left(k_{yn}^{'(2)}\, d \right)
\end{equation}
where $d$ is the separation between the moving grating and the metallic plate. The rest of the terms in the definition of $Y'_\mathrm{eq}$ retain the same meaning as in the main text. Furthermore, the $n=0$ wave admittance of the metal-backed output medium in the laboratory frame, $Y_{0}^{(2)}$, has to be similarly updated in  the reflection coefficient $R$ formula (Eq.\eqref{R}). Note that these corrections already include the effect of the higher-order modes.

\bibliography{apssamp}

\end{document}